\documentclass[aps,preprint,superscriptaddress,showpacs,preprintnumbers,amsmath,amssymb]{revtex4}

\usepackage{graphicx}
\usepackage{dcolumn}
\usepackage{bm}
\usepackage{float,color}
\usepackage{CJK}

\begin{document}
\begin{CJK*}{GBK}{song}

\title{Pairing strength on the nuclear size in relativistic continuum Hartree-Bogoliubov theory}

\author{Ying Chen }
\affiliation{State Key Laboratory of Nuclear Physics and Technology, School of Physics, Peking University, Beijing 100871, China}

\author{Peter Ring}
\affiliation{State Key Laboratory of Nuclear Physics and Technology, School of Physics, Peking University, Beijing 100871, China}
\affiliation{Physik-Department der Technischen Universit\"at M\"unchen, D-85748 Garching, Germany}

\author{Jie Meng \footnote{Email: mengj@pku.edu.cn}}
\affiliation{State Key Laboratory of Nuclear Physics and Technology, School of Physics, Peking University, Beijing 100871, China}
\affiliation{School of Physics and Nuclear Energy Engineering, Beihang University, Beijing 100191, China}
\affiliation{Department of Physics, University of Stellenbosch, Stellenbosch, South Africa}

\date{\today}

\begin{abstract}
The influence of pairing correlations on the nuclear size and in particular on the formation
of nuclear halos is studied in the framework of relativistic continuum Hartree-Bogoliubov (RCHB)
theory. It turns out that the contributions from the weakly-bound orbits with low orbital angular
momenta $l$ play an important role. As an example, we investigate the neutron-rich Mg isotopes as
a function of the pairing strength in situations, where the neuron Fermi surface are below, between
and above two weakly-bound $2p$ levels. We find that the size of the pairing
correlations has a two-fold influence on the density distribution of the neutrons and therefore
on the total nuclear size. First it can change the root mean square (rms) radius of the individual
weakly-bound orbits and second it can enhance the occupation probabilities of these orbits in
the nuclear system. On one side increasing pairing correlations reduce the rms radii of the orbits
with small orbital angular momenta close to the continuum limit (pairing anti-halo effect), on
the other side they also can lead to an enhanced occupation of low-$l$ orbits above the Fermi
surface producing in this way a strong increase of the total radius of the nuclear system.
As a consequence, a nuclear halo can form even in cases, where non of the individual low-$l$
orbits is very close to the continuum. This leads to the fact that compared with well-bound
nuclei, the impact of the pairing strength on the nuclear size is more pronounced in weakly-bound
nuclei than in well-bound systems.
\end{abstract}

\pacs{21.10.Gv, 21.60.Jz, 27.40.+z}

\maketitle


Since the experimental discovery of the anomalously large nuclear radius in
$^{11}$Li~\cite{Tanihata1985Phys.Rev.Lett.2676}, which is far beyond the conventional
$A^{1/3}$ law of the mass dependence, considerable experimental and theoretical
efforts~\cite{Bertsch1991_APNY209-327} have been undertaken to understand this
interesting phenomenon of the formation of a neutron halo in exotic nuclei close
to the neutron drip-line.

In the mean-field descriptions of nuclear halos, one interpretation is the pairing effect~\cite{Meng1996Phys.Rev.Lett.3963,Meng1998Phys.Rev.Lett.460,Meng1998PRC,Meng2002Phys.Rev.C041302}.
With the relativistic continuum Hartree-Bogoliubov (RCHB) theory~\cite{Meng1998Nucl.Phys.A3},
the first self-consistent microscopic description of the neutron halo in $^{11}$Li has been
presented in Ref.~\cite{Meng1996Phys.Rev.Lett.3963}, where the halo is reproduced by scattering
Cooper pairs from the $1p_{1/2}$ level to the $2s_{1/2}$ level in the continuum, demonstrating
that the coupling between the bound states and the continuum is of great importance to account
for the formation of a halo. Subsequently, another interesting phenomenon, a giant halo, in
exotic Ca and Zr isotopes are predicted by RCHB theory, which are formed by more than two
neutrons scattered as Cooper pairs to the continuum region~\cite{Meng1998Phys.Rev.Lett.460,
Meng1998Nucl.Phys.A3,Meng2002Phys.Rev.C041302,Zhang2002CPL,zhang2003SIC}.
In recent years, in order to study the halo phenomena in deformed nuclei, a deformed relativistic
Hartree-Bogoliubov (RHB) theory in continuum has been developed~\cite{Zhou2003Phys.Rev.C034323,
Zhou2006AIPConf.Proc.90,Li2012PhysRevC.85.024312,Li2012CPL,Chen2012PhysRevC.85.067301},
and a shape decoupling between the core and halo has been
predicted~\cite{Zhou2003Phys.Rev.C034323,Li2012PhysRevC.85.024312}.

On the other hand, in Ref.~\cite{Bennaceur2000PLB}, the asymptotic Hartree-Fock (HF)
and Hartree-Fock-Bogoliubov (HFB) densities characterized by $l=0$ orbits are compared
and it has been emphasized that pairing correlations reduce the nuclear size and an extreme
halo with infinite radius cannot be formed in nuclear systems. The mean square radius
calculated by the asymptotic HF density is
\begin{equation}
    \langle r^2\rangle_{\rm{HF}} \propto \frac{\hbar^2}{2m|\varepsilon_k|},
\end{equation}
which will diverge in the limit of vanishing separation energy, the single particle energy
$\varepsilon_k \rightarrow 0$. In contrast, the mean square radius calculated by the asymptotic
HFB density is
\begin{equation}
    \langle r^2\rangle_{\rm{HFB}} \propto \frac{\hbar^2}{2m\Delta_k},
\end{equation}
which will not diverge if the the pairing gap $\Delta_k$ stays finite in the limit of small
separation energy. This effect that additional pairing correlation acts against a development of an
infinite radius is then concluded as ``pairing anti-halo effect"~\cite{Bennaceur2000PLB}.

However, for realistic nuclei further points have to be taken into account. First, the continuum
cannot be included in HF calculations, therefore the above discussions are confined to the case
that the valence nucleons occupy the weakly-bound orbits below the continuum threshold, which is
also an early interpretation for the nuclear halo~\cite{Bertsch1989_PRC39-1154,Sagawa1992_PLB286-7,ZHU-ZY1994_PLB328-1}.
Second, the limiting condition $\varepsilon_k \rightarrow 0$, and estimation of the nuclear rms
radius calculated by the asymptotic wave functions instead of the real ones correspond to an extremely
ideal situation, which is actually difficult to reach in the real nucleus.

In the present investigation, we study the influence of pairing correlations on the nuclear size,
in particular on the formation of a possible halo. We use RCHB theory, where the continuum is taken
in to account properly, and consider the neutron-rich nuclei with the neutron Fermi surface below,
between and above two weakly-bound $2p$ levels.

  \begin{figure}[tbh]
      \centering
      \includegraphics[width=8cm]{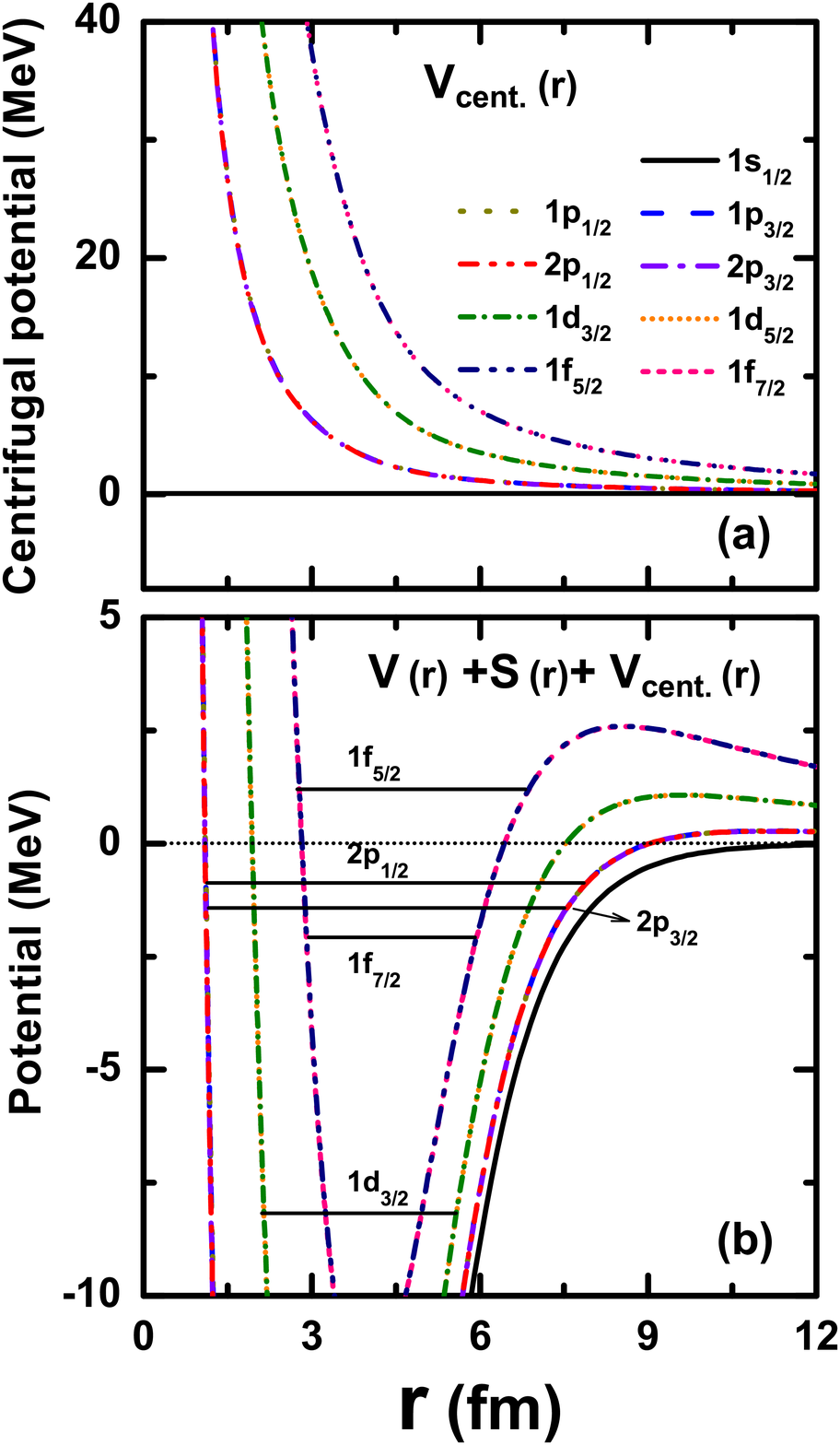}
      \caption{(Color online)  (a) Centrifugal potentials $V_{\rm{cent.}}(r)$ and (b) potentials $V(r)+S(r)+V_{\rm{cent.}}(r)$
               of Woods-Saxon shape for the single neutron levels with the orbital quantum number $l=0,~1,~2,~3$. We also
               show levels in the vicinity of the continuum threshold.
               }\label{fig1}
  \end{figure}

The RHB equations~\cite{Kucharek1991_ZPA339-23} for the nucleons read as
\begin{eqnarray}
    \left(\begin{array}{cc}
        h_D-\lambda & \Delta\\
        -\Delta^*& -h^*_D+\lambda
    \end{array}\right)
    \left(\begin{array}{c}
        U_k\\
        V_k
    \end{array}\right)
    =E_k\left(\begin{array}{c}
        U_k\\
        V_k
    \end{array}\right)\label{eq1}
\end{eqnarray}
with the quasiparticle energies $E_k$, the Fermi surface $\lambda$, and the Dirac Hamiltonian
\begin{equation}
    h_D=\mathbf{\alpha}\cdot\mathbf{p}+ \mathbf{\beta}M + \Sigma (\mathbf{r}),
\end{equation}
where the mean-field
\begin{equation}
    \Sigma(\mathbf{r}) = V(\mathbf{r})+\mathbf{\beta}S(\mathbf{r}) \label{eq2}
\end{equation}
contains scalar and vector potentials $S(\mathbf{r})$ and $V(\mathbf{r})$. For the following
investigations, we use the relativistic energy density functional PK1, which has been developed
in Ref.~\cite{Long2004Phys.Rev.C034319}. However, this is not essential, because it is known from the literature~\cite{Meng1996Phys.Rev.Lett.3963,Meng1997_ZPA358-123,Poschl1997_PRL79-3841,
Meng1998Phys.Rev.Lett.460,Meng1998_PLB419-1,Meng1998Nucl.Phys.A3,Vretenar1998_PRC57-1060,
Stoitsov1998_PRC58-2086,Meng1999_NPA654-702c,Meng2002Phys.Rev.C041302,Zhang2002CPL,zhang2003SIC,
Meng2003_NPA722-366c,Long2010_PRC81-031302,ZHOU-SG2010_PRC82-011301,
ZHOU-SG2011_JPCS312-092067} that other successful density functionals show a similar behavior.

Pairing correlations are introduced by a density dependent effective pairing force of zero range~\cite{Bertsch1991_APNY209-327}:
\begin{eqnarray}
    V^{pp}(\mathbf{r_1},\mathbf{r_2})&=&\frac{v_0}{2}(1-P^{\sigma})\delta(\mathbf{r_1}-\mathbf{r_2})(1-\frac{\rho(\mathbf{r_1})}{\rho_{\rm{sat}}}),
\end{eqnarray}
where $\rho_{\rm{sat}}=0.152$ fm$^{-3}$ is the saturation density. It leads to a local pairing potential
\begin{equation}
    \Delta(\mathbf{r})= v_0(1-\frac{\rho(\mathbf{r})}{\rho_{\rm{sat}}})\kappa (\mathbf{r})\label{eq3}
\end{equation}
with the pairing tensor \begin{equation}
    \kappa(\mathbf{r})= \sum_{k>0} U_k(\mathbf{r})V_k(\mathbf{r}).\label{eq4}
\end{equation}

In order to describe the continuum and its coupling to the bound states properly,
the RCHB equations are solved in coordinate space in a finite box by the method discussed in Refs.~\cite{Meng1998Nucl.Phys.A3,Meng2006Prog.Part.Nucl.Phys.470}. For a large box, this
method presents a reliable solution of these equations~\cite{Grasso2001_PRC64-064321,zhang2011PRC}.
In the following calculations, the RCHB equations are solved in a box with the size $R=20$
fm and a step size of $0.1$ fm and it is checked that for these values of the grid suitable
convergence is achieved for all the results presented here. The influence of pairing correlations
on the halo is investigated by using a variable strength $v_0$ of the pairing interaction.
The contribution from the continuum is restricted within a cutoff energy $E_{\rm{cut}}\sim 120$ MeV.

Considering spherical symmetry, the mean-field $\Sigma(\bm r)$ in Eq.~(\ref{eq2}) and the pairing
potential $\Delta(\bm r)$ in Eq.~(\ref{eq4}) depend only on the the radial coordinate. In order to
clarify the situation, we first consider potentials of Woods-Saxon (WS) shape:
\begin{equation}
     \Sigma_\pm(r)= \frac{V_{\rm{0}}}{1+e^{(r-R)/a}}\label{eq:WS}
\end{equation}
with the parameters $V_0=-51.32$ MeV, $R=4.42$ fm, $a=1.00$ fm for $\Sigma_+(r)=V(r)+S(r)$
and $V_0=576.34$ MeV, $R=4.10$ fm, $a=0.65$ fm for $\Sigma_-(r)=V(r)-S(r)$. These values have
been obtained by a fit to the corresponding neutron potentials resulting from a self-consistent
calculation in the neutron-rich nucleus $^{42}$Mg as discussed below. Here the weakly-bound $2p$ levels
are close to the continuum threshold (see Fig.~\ref{fig1}).

In the mean-field approximation without pairing correlations, i.e. for $\Delta (\bm r)=0$, the RCHB
equation can be reduced to a Dirac equation. In the spherical case, starting from the radial
Dirac equation, one can obtain a Schr\"{o}dinger-like equation for the upper components of the
wave function with the potential $V(r)+S(r)+V_{\rm{cent.}}(r)$~\cite{Zhang-Ying2010_IJMPE19-55}.
The centrifugal potential $V_{\rm{cent.}}(r)$ is represented as
\begin{equation}
    V_{\rm{cent.}}(r)=\frac{\hbar^2}{2M_{+}(r)}\frac{l(l+1)}{r^2}
\end{equation}
with
\begin{equation}
    2M_+(r)=2M+S(r)-V(r)+\varepsilon_k,
\end{equation}
where $\varepsilon_k$ is the corresponding single particle energy .

Fig.~\ref{fig1} shows the centrifugal potentials $V_{\rm{cent.}}(r)$ (panel (a)) and the
potentials $\Sigma_+(r)+V_{\rm{cent.}}(r)$ (panel (b)) for the single neutron levels with the
orbital quantum number $l=0,~1,~2,~3$. Due to the fact that for the levels in the neighborhood
of the continuum limit with $\varepsilon_k\ll S-V$, no observable differences show up for the
states with the same $l$, neither in the centrifugal potentials $V_{\rm{cent.}}(r)$ nor in the
potentials $\Sigma (r)+V_{\rm{cent.}}(r)$. For increasing $l$-values, the centrifugal potential
$V_{\rm{cent.}}(r)$ becomes larger, and thereby the potential $V(r)+S(r)+V_{\rm{cent.}}(r)$
becomes narrower with increasing orbital angular momentum. For the $p,~d,~f$ levels, the barriers
of the potentials $\Sigma_+(r)+V_{\rm{cent.}}(r)$ in the region  $r> 7$ fm are about
$0.25,~1.00,~2.50$ MeV, respectively.

After the solution of the Dirac equation with the WS potential, one obtains the single neutron
levels. In panel (b) of Fig.~\ref{fig1}, we show the corresponding bound states
$1d_{3/2},~1f_{7/2},~2p_{3/2},~2p_{1/2}$ and a resonance state $1f_{5/2}$ in the neighborhood of
the continuum threshold. The gaps between the $1d_{3/2}$ and $1f_{7/2}$ orbits and between the $1f_{7/2}$
and $2p_{3/2}$ orbits are $6.09$ and $0.67$ MeV, respectively. These gaps correspond to the traditional
magic numbers $N=20$ and $N=28$. It is evident that the shell gap at $N=28$ is reduced considerably as
compared to the one at $N=20$. The traditional magic number $N=28$ has basically disappeared. The reason
for this is the fact that the depth and width of the WS potential used here is determined by the
self-consistent neutron potential in the nucleus $^{42}$Mg. Since $^{42}$Mg, containing $30$ neutrons,
is extremely neutron-rich, the self-consistent calculation produces a highly diffuse neutron potential,
and thereby the $2p_{3/2}$ level is shifted downwards. As a consequence, the magicity at $N=28$ is lost.

In the first part of our investigations, in order to concentrate on the influence of pairing correlations on
the nuclear size, we will keep the WS potentials fixed at the values adjusted to the neutron-rich nucleus $^{42}$Mg,
i.e. at the values shown in Fig.~\ref{fig1}.

\begin{figure}[tbh]
    \centering
    \includegraphics[width=8cm]{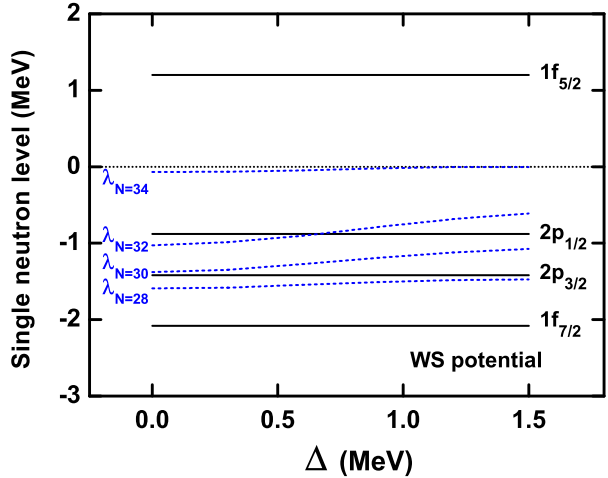}
    \caption{(Color online) Single neutron levels $2p$ and $1f$ in the canonical basis as a function of
             average pairing gap $\Delta$ in Eq.~(\ref{eq12}). The Fermi surfaces for the neutron numbers
             $N=28,~30,~32,~34$ are plotted as dashed lines.
             }\label{fig2}
\end{figure}

It is known that the weakly-bound states and the resonance states in the continuum play a crucial role in the
formation of a nuclear halo~\cite{Meng1996Phys.Rev.Lett.3963,Meng1998Phys.Rev.Lett.460}. Therefore
we solve the corresponding RCHB equations~(\ref{eq1}) with neutron number
$N=34$ for various pairing strength $v_0$ and show in Fig.~\ref{fig2} the single neutron levels
$2p_{3/2},~2p_{1/2},~1f_{7/2}$ and $1f_{5/2}$ in canonical basis~\cite{Ring1980} as a function of
the average pairing gap,
\begin{equation}
    \Delta = \frac{1}{N}\sum_{nlj} (2j+1) \Delta^{(lj)}_{n} v_{nlj}^2,\label{eq12}
\end{equation}
where $\Delta^{(lj)}_{n}$ are the diagonal matrix elements of the pairing field in the canonical
basis and $v_{nlj}^2$ are the corresponding occupation probabilities. The canonical basis is calculated
after a self-consistent solution of the RCHB equations~(\ref{eq1}) and therefore the energy levels
$\varepsilon_{nlj}=h^{(lj)}_{nn}$ depend in principle on $\Delta$. However, since this solution was
obtained for fixed WS potentials, we find in Fig.~\ref{fig2} that these states remain almost
unchanged with the increasing pairing correlations.

In order to illustrate the influence of pairing correlations on the nuclear size, we choose the four
neutron-rich isotopes with $N=28,~30,~32$, and $34$ as examples. For these four cases, the neutron Fermi
surface (shown as dashed line in Fig.~\ref{fig2}) is located just below, between and above the two
weakly-bound $2p$ levels. As the pairing strength increases, more neutrons are scattered from occupied
levels in the Fermi sea to empty levels above the Fermi sea, and therefore the Fermi surface is raising
at the same time.

\begin{figure*}[tbh]
     \centering
     \includegraphics[width=14cm]{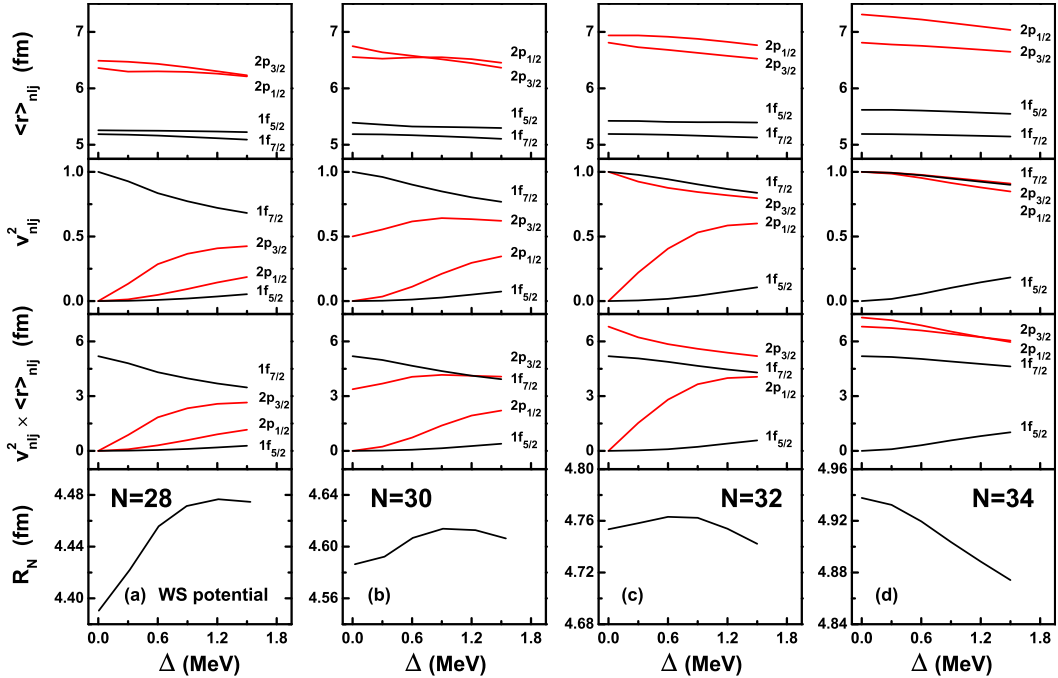}
     \caption{(Color online) Rms radii $\langle r\rangle_{nlj}$, occupation probabilities $v^2_{nlj}$,
              contributions $v^2_{nlj}\times\langle r\rangle_{nlj}$ to the neutron rms radii for the $2p$ and $1f$
              orbits, and the total neutron rms radii $R_{N}$ for $N=28,~30,~32,~34$ as a function of average pairing
              gap $\Delta$.
              }\label{fig3}
\end{figure*}

In Fig.~\ref{fig3}, the rms radii $\langle r\rangle_{nlj}$, occupation probabilities $v^2_{nlj}$,
 corresponding contributions $v^2_{nlj}\times \langle r\rangle_{nlj}$ to the neutron rms radius for
$2p$ and $1f$ orbits, and total neutron rms radii $R_{N}$ are plotted as a function of the average pairing
gap for the different neutron-rich isotopes with $N=28,~30,~32$ and $34$. The rms radius
$\langle r\rangle_{nlj}$ for one single neutron level denoted by quantum number $nlj$ is defined as
\begin{equation}
  \langle r \rangle_{nlj} =\left(\frac{\int 4\pi r^4 \rho_{nlj}(r)dr}{\int 4\pi r^2 \rho_{nlj}dr}\right)^{1/2},
\end{equation}
where $\rho_{nlj}(r)=|\psi_{nlj}(r)|^2$ represents the probability density of each orbit with the wave
function $\psi_{nlj}$ in canonical basis~\cite{zhang2003SIC}. In the upper panels, it is clearly seen that
the rms radius of the $2p$ orbits are much larger than those of the $1f$ orbits, due to the lower
centrifugal barrier. With increasing pairing correlations, the rms radius decreases for the $2p$ and
the $1f$ levels, especially for the $2p$ states. The reason for this decrease is the so-called
``pairing anti-halo effect" discussed in Ref.~\cite{Bennaceur2000PLB}. However, this effect concerns
the radii $\langle r\rangle_{nlj}$ of the individual orbits. On the other hand, the total neutron radius
of the nucleus is given by
\begin{equation}
      R_N= \sum_{nlj} (2j+1)\,\langle r\rangle_{nlj}\,v^2_{nlj}.
\end{equation}
It is determined by the rms radii of the orbits and the corresponding occupation probabilities,
which depend strongly on the pairing correlations.

For $N=28$ in panel (a), the neutron Fermi surface is just below the two weakly-bound $2p$ levels
shown in Fig.~\ref{fig2}. Without pairing correlations, the occupation probability is $1.0$ for the
$1f_{7/2}$ orbit, and it vanishes for the $2p$ and the $1f_{5/2}$ orbits. As the pairing strength
increases, the neutrons on $1f_{7/2}$ orbit are scattered to $2p$ and $1f_{5/2}$ orbits which
have much larger rms radii. Therefore, the contributions to the total neutron rms radius from $2p$
and $1f_{5/2}$ states grow more than the contribution from $1f_{7/2}$ state decreases. As a results,
the total neutron rms radius increases monotonically, where the contributions from $2p$ orbits play
a dominant role.

Adding two more neutrons, i.e. for $N=30$ in panel (b), for zero pairing, the $2p_{1/2}$ and the $1f_{5/2}$
orbits are empty, while the $1f_{7/2}$ orbit is fully occupied and the $2p_{3/2}$ orbit is half occupied.
With increasing pairing strength, the neutrons on $1f_{7/2}$ orbit are scattered to the $2p$ orbits
which contribute strongly to the neutron rms radius. Therefore, the total neutron rms radius $R_N$ increases
with the growing pairing strength up to $\Delta=0.9$ MeV. Increasing the pairing strength further, the
neutrons are continued to be scattered from the $1f_{7/2}$ to the $2p_{1/2}$ state, while the neutron
number in the $2p_{3/2}$ orbit decreases slightly and the resonance state $1f_{5/2}$ begins to be occupied.
Together with the monotone decreasing rms radii of the individual orbits, the increase resulting from
the $2p_{1/2}$ and $1f_{5/2}$ states is smaller than the decreases of the contributions from the
$2p_{3/2}$ and $1f_{7/2}$ states. Therefore, for very strong pairing correlations, the total neutron
rms radius finally decreases.

  \begin{figure}[tbh]
      \centering
      \includegraphics[width=8cm]{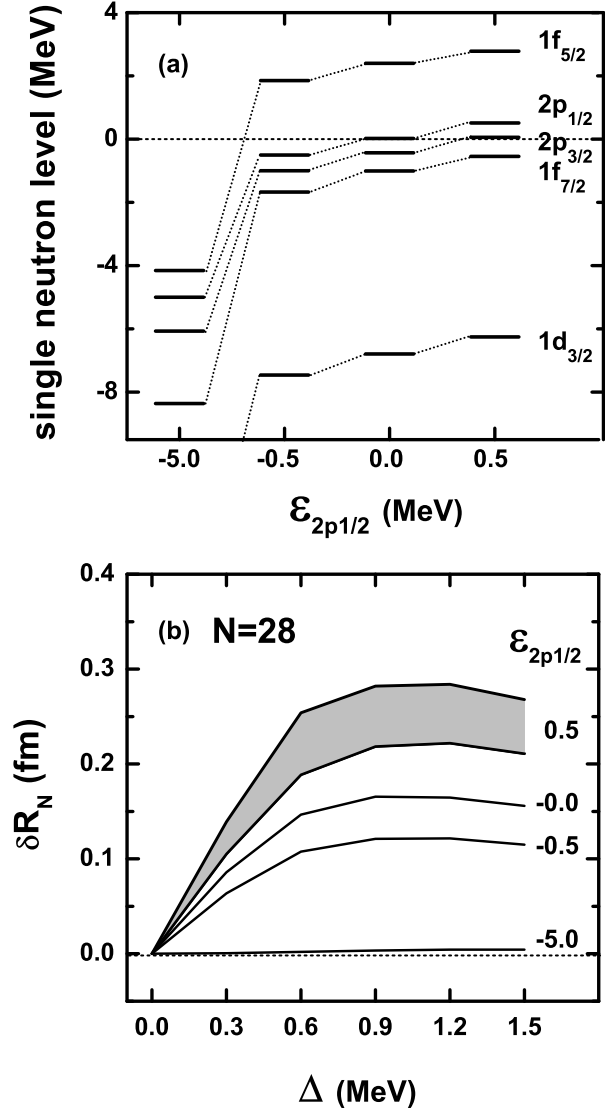}
      \caption{(Color online)(a) Single neutron levels $1d_{3/2},~1f_{7/2},~2p_{3/2},~2p_{1/2},~1f_{5/2}$ in depth-variant
                WS potentials where $\varepsilon_{2p_{1/2}}=-5.0,~-0.5,~0.0,~0.5$ MeV. (b) Changes of the total
                neutron rms radii with respect to those of zero pairing $\delta R_{N}=R_N(\Delta)-R_N(\Delta=0.0)$ for each
                $\varepsilon_{2p_{1/2}}$ case as a function of average pairing gap $\Delta$. The shadow area represents the
                results calculated with different box sizes in the WS potential with $\varepsilon_{2p_{1/2}}=0.5$ MeV .
                }\label{fig4}
  \end{figure}

\begin{figure*}[tbh]
      \centering
      \includegraphics[width=10cm]{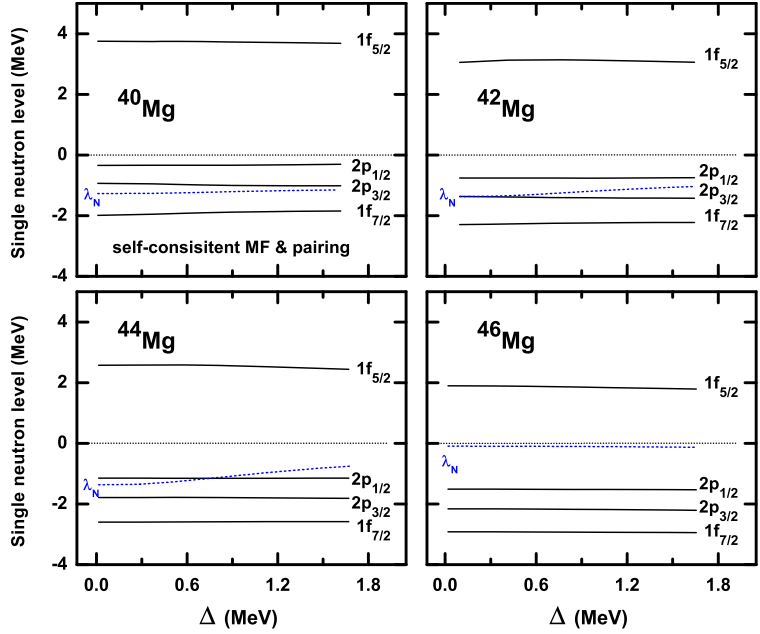}
      \caption{(Color online) Same as Fig.~\ref{fig2}, but both mean-field and pairing potential are
               calculated self-consistently.
               }\label{fig5}
\end{figure*}

With two more neutrons, i.e. for $N=32$ in panel (c), without pairing correlation, the $1f_{7/2}$ and
$2p_{3/2}$ orbits are fully occupied, while the $2p_{1/2}$ and $1f_{5/2}$ orbits are empty. As the
pairing strength increases, the neutrons on the $1f_{7/2}$ and $2p_{3/2}$ orbits are scattered to the
$2p_{1/2}$ orbit, whose rms radius is larger than the ones of $1f_{7/2}$ and $2p_{3/2}$ levels. The
contribution of the $2p_{1/2}$ orbits cause an increase in the total neutron rms radius up to a pairing
gap $\Delta =0.9$ MeV.  As the pairing strength continues to grow, the neutrons on the $2p_{3/2}$ state
begins to be scattered to the $1f_{5/2}$ state which provides a smaller contribution than the $2p_{3/2}$
state, while the occupation probability remains almost constant for the $1p_{1/2}$ state. Together with
the decreasing rms radius of the individual orbits, the total neutron rms radius finally decreases for very
large pairing.

\begin{figure*}[tbh]
      \centering
      \includegraphics[width=14cm]{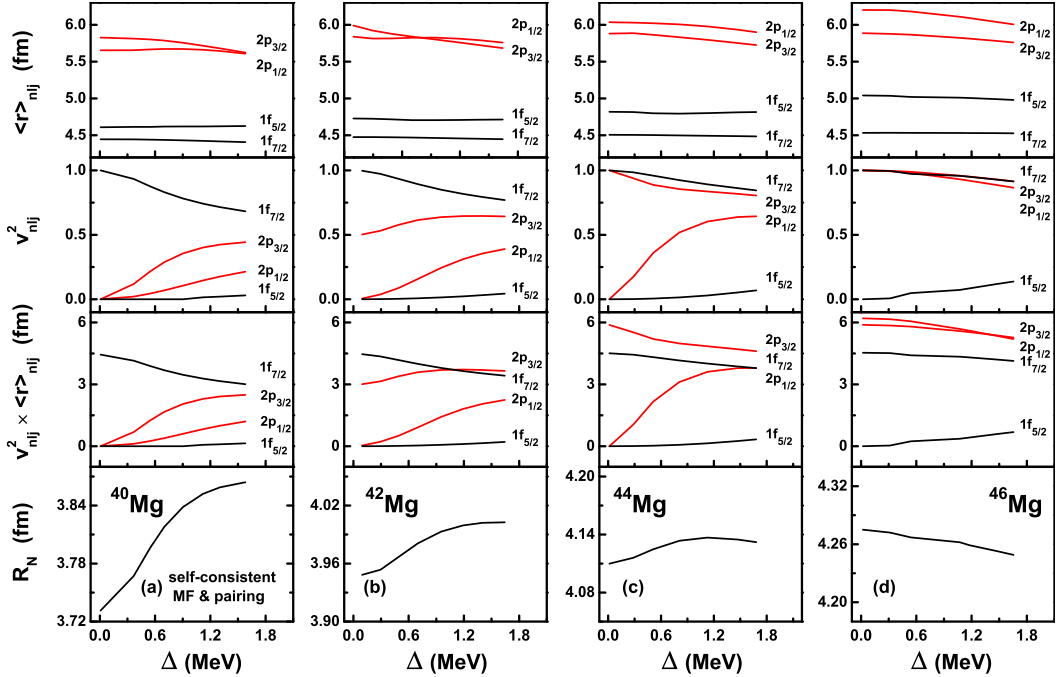}
      \caption{(Color online) Same as Fig.~\ref{fig3},  but both mean-field and pairing potential
                are calculated self-consistently.
                }\label{fig6}
\end{figure*}

Considering the two weakly-bound $2p$ levels fully occupied, i.e. for $N=34$ in panel (d), for zero pairing,
the neutron Fermi surface is just above the weakly-bound $2p$ levels. With increasing pairing strength,
the neutrons on the $1f_{7/2}$ and $2p$ orbits are scattered to the $1f_{5/2}$ orbit with a much smaller
rms radius than the $2p$ orbits. The contributions to the total neutron rms radius from the $1f_{7/2}$
and $2p$ orbits decrease more than the increase of the contribution from $1f_{5/2}$ orbit, and therefore
the total neutron rms radius finally decreases monotonically. It should be emphasized that here the two
effects, decreasing rms radius of the individual orbits and changes of the occupation probabilities, act
in the same direction.  The decrease of the total neutron radius $R_N$ with increasing pairing correlations
comes not only from the decreasing rms radii of the individual orbits but also from the change of the
occupation probabilities. As the pairing gap increases from $0$ to $1.5$ MeV, the occupation probabilities
of the $2p_{1/2}$ and $2p_{3/2}$ states decrease by $15.2$\% and $9.10$\%, respectively. The corresponding
decrease of the total radius is much larger than that caused by the decrease of the rms radii of the
individual orbits of about $3.76$\% and $2.38$\%. Therefore the decrease of the contributions of the $2p$ orbits
is mainly caused by the decreasing occupation probabilities.

It is evident that pairing correlations can change the rms radii of the individual weakly-bound orbits and
simultaneously the corresponding occupation probabilities. As a result, they  have a strong influence the
nuclear rms radius and the total nuclear size. Furthermore, the contributions of the weakly-bound $2p$ orbits
to the nuclear rms radius  play an important role.

In the next step, we investigate the impact of pairing correlations on the nuclear size when the $2p$ orbits
are well-bound, weakly-bound, around the continuum threshold, and finally in the continuum. As an example, we consider
$N=28$ and change the depth of the WS potential in such a way that the position of the $\varepsilon_{2p_{1/2}}$
changes. Panel (a) shows the single neutron levels $1d_{3/2},~1f_{7/2},~2p_{3/2},~2p_{1/2},~1f_{5/2}$ in WS potentials of
varying depth such that $\varepsilon_{2p_{1/2}}=-5.0,~-0.5,~0.0,~+0.5$ MeV. In panel (b), for each case of
$\varepsilon_{2p_{1/2}}$, the change of the total neutron rms radius with respect to that of zero pairing
$\delta R_{N}=R_N(\Delta)-R_N(\Delta=0.0)$ is plotted as a function of the pairing gap.

It is found that with the pairing gap increasing from $0.0$ to $1.5$ MeV, the neutron rms radius $R_N(\Delta)$
increases only slightly by $\sim0.004$ fm for the case $\varepsilon_{2p_{1/2}}=-5.0$ MeV. This corresponds to
a well bound nucleus. On the other side, we observe significant increases by more than $0.1$ fm for the cases
$\varepsilon_{2p_{1/2}}=-0.5,~0.0,~+0.5$ MeV which correspond to the weakly-bound nuclei. Here it has to be
noticed that when $\varepsilon_{2p_{1/2}}=+0.5$ MeV and $\varepsilon_{2p_{3/2}}=0.05$ MeV in the continuum,
the system is basically not bound and the solution depends on the box size we used. For an illustration, by
changing the box size from $20$ to $25$ fm, we find solutions of the shadowed region of panel (b). It can be
seen that the neutron rms radius calculated with the $25$ fm box size is about $0.05$ fm larger than the one
obtained with $20$ fm box size when $\Delta=1.5$ MeV. Nevertheless, we can conclude that the influence of
pairing correlations on the nuclear size is much more significant for weakly-bound nuclear than for well-bound ones.

In the calculations discussed so far, we have kept the mean-fields fixed in the form of spherical WS potentials,
only changing the pairing strength. We investigate the influence of pairing correlations on the nuclear size for
$N=28,~30,~32,~34$, where the Fermi surfaces are located below, between and above the two weakly-bound $2p$ levels.
Here, with increasing pairing strength, the single neutron levels stay almost a constant. In the next step, we
investigate what happens if both the mean-fields and the pairing potential is changing in a consistent way.
We consider the neutron-rich nuclei $^{40,~42,~44,~46}$Mg as examples and use the effective interaction PK1
developed in Ref.~\cite{Long2004Phys.Rev.C034319} in fully self-consistent solutions of the RCHB equations~(\ref{eq1}).

The single neutron levels $2p$ and $1f$ in the canonical basis are shown as a function of average pairing gap for the
nuclei $^{40,~42,~44,~46}$Mg in Fig.~\ref{fig5}. It can be seen that for each nucleus the weakly-bound $1f_{7/2}$,
$2p_{1/2}$ and $2p_{3/2}$ levels are around the Fermi surfaces, and that the resonance state $1f_{5/2}$ is above
the continuum threshold. With increasing neutron number, the neutron potential becomes broader and more diffuse.
Thereby the $2p$ and $1f$ states drop down from $^{40}$Mg to $^{46}$Mg. Furthermore, with increasing pairing strength,
the $2p$ and $1f$ states show a slight decrease for each nucleus.

In Fig.~\ref{fig6}, we show, similar to Fig.~\ref{fig3}, the results of self-consistent calculations for the rms radii
and occupation probabilities of the individual single neutron levels $2p$ and $1f$ and for the the total neutron rms radii
$R_{N}$ as a function of average pairing gap in the nuclei $^{40,~42,~44,~46}$Mg. We observe very similar tendencies in the
changes of the neutron rms radius with increasing pairing strength as it has been found for fixed WS potentials in
Fig.~\ref{fig3}, but with different amplitudes, especially for the nuclei $^{40}$Mg and $^{46}$Mg. For $^{40}$Mg,
as it contains only $28$ neutrons, the neutron potential is narrower and shallower than the WS potential adjusted
the self-consistent neutron potential of $^{42}$Mg. As compared to Fig.~\ref{fig2}, the single neutron levels raise
up in Fig.~\ref{fig5}, especially the $1f_{5/2}$ state. The energy gap between $2p_{1/2}$ and $1f_{5/2}$ states
is about $4.0$ MeV and therefore it is more difficult to scatter neutrons from the $2p$ orbits to the $1f_{5/2}$
orbit. As a result, the total neutron rms radius of $^{40}$Mg increases more. For $^{46}$Mg, the self-consistent
neutron potential is broader and more diffuse than the WS potential adjusted to $^{42}$Mg and therefore the $2p$
and $1f$ orbits come down considerably. Their rms radius is significantly reduced, especially for the $2p$ levels.
Therefore, with increasing pairing strength, even if neutrons are scattered from the $2p$ to the $1f_{5/2}$ levels,
the amplitudes of the change in the neutron rms radius $R_N$ is not so significant due to the smaller differences
between the rms radii of the $2p$ and the $1f_{5/2}$ states. Nevertheless, it should be emphasized that qualitatively
the effect of pairing correlations on the nuclear size agrees in the fully self-consistent calculations with that found
for fixed WS potentials.

Summarizing, the relativistic continuum Hartree-Bogoliubov (RCHB) theory is used to study the influence of pairing
correlations on the nuclear size for neutron-rich nuclei. It is found that the weakly-bound orbits with low orbital
angular momenta and small centrifugal barriers as for instance the $2p$ levels in our investigation have an important
influence on the total neutron radius and on the development of a halo. Investigating cases where the neutron Fermi
surface is below, between and above the weakly-bound $2p$ levels, it is found that pairing correlations have a
two-fold influence on the total nuclear rms radius and the nuclear size. First, they can change the rms radii of individual
weekly bound orbits and, second, they can change their occupation probabilities. With increasing pairing strength, the
individual rms radii are reduced (pairing anti-halo effect). This effect depends on the distance from the continuum limit.
More important is in most of the cases the change of the occupation probabilities. If the low-$l$ orbits are close to
the continuum limit, they are in many cases not occupied for small or vanishing pairing correlations. Increasing pairing
leads to increasing occupation numbers for these important orbits and consequently to considerable changes of the total
neutron radius and the size of the nucleus. Only in cases, where these orbits are below the Fermi surface and still close
to the continuum limit, increasing pairing correlations change their occupations and reduce the nuclear size. This effect
is often more important than the change of the individual rms radii of these orbits.

In the present investigations, the strength of the pairing correlations was an external variable parameter. Of course,
in realistic nuclei the strength of pairing depends on the level density in the vicinity of the Fermi surface. Therefore,
even orbits with high $l$-values and large centrifugal barriers can contribute indirectly to changes of the nuclear size.
If they come close to the Fermi surface they enhance pairing correlations influencing at the same time the occupation of
low-$l$ orbits close to the continuum limit and therefore the nuclear size.

\section*{ACKNOWLEDGMENTS}
This work was partially supported by the Major State 973 Program 2007CB815000, National Natural Science Foundation of
China under Grant Nos. 11175002, and the Research Fund for the Doctoral Program of Higher Education under Grant No. 20110001110087.


\bibliographystyle{apsrev}


\begin{thebibliography}{35}
\expandafter\ifx\csname natexlab\endcsname\relax\def\natexlab#1{#1}\fi
\expandafter\ifx\csname bibnamefont\endcsname\relax
  \def\bibnamefont#1{#1}\fi
\expandafter\ifx\csname bibfnamefont\endcsname\relax
  \def\bibfnamefont#1{#1}\fi
\expandafter\ifx\csname citenamefont\endcsname\relax
  \def\citenamefont#1{#1}\fi
\expandafter\ifx\csname url\endcsname\relax
  \def\url#1{\texttt{#1}}\fi
\expandafter\ifx\csname urlprefix\endcsname\relax\def\urlprefix{URL }\fi
\providecommand{\bibinfo}[2]{#2}
\providecommand{\eprint}[2][]{\url{#2}}

\bibitem[{\citenamefont{Tanihata et~al.}(1985)\citenamefont{Tanihata, Hamagaki,
  Hashimoto, Shida, Yoshikawa, Sugimoto, Yamakawa, Kobayashi, and
  Takahashi}}]{Tanihata1985Phys.Rev.Lett.2676}
\bibinfo{author}{\bibfnamefont{I.}~\bibnamefont{Tanihata}},
  \bibinfo{author}{\bibfnamefont{H.}~\bibnamefont{Hamagaki}},
  \bibinfo{author}{\bibfnamefont{O.}~\bibnamefont{Hashimoto}},
  \bibinfo{author}{\bibfnamefont{Y.}~\bibnamefont{Shida}},
  \bibinfo{author}{\bibfnamefont{N.}~\bibnamefont{Yoshikawa}},
  \bibinfo{author}{\bibfnamefont{K.}~\bibnamefont{Sugimoto}},
  \bibinfo{author}{\bibfnamefont{O.}~\bibnamefont{Yamakawa}},
  \bibinfo{author}{\bibfnamefont{T.}~\bibnamefont{Kobayashi}},
  \bibnamefont{and}
  \bibinfo{author}{\bibfnamefont{N.}~\bibnamefont{Takahashi}},
  \bibinfo{journal}{Phys. Rev. Lett.} \textbf{\bibinfo{volume}{55}},
  \bibinfo{pages}{2676} (\bibinfo{year}{1985}).

\bibitem[{\citenamefont{Bertsch and Esbensen}(1991)}]{Bertsch1991_APNY209-327}
\bibinfo{author}{\bibfnamefont{G.~F.} \bibnamefont{Bertsch}} \bibnamefont{and}
  \bibinfo{author}{\bibfnamefont{H.}~\bibnamefont{Esbensen}},
  \bibinfo{journal}{Ann. Phys.} \textbf{\bibinfo{volume}{209}},
  \bibinfo{pages}{327 } (\bibinfo{year}{1991}).

\bibitem[{\citenamefont{Meng and Ring}(1996)}]{Meng1996Phys.Rev.Lett.3963}
\bibinfo{author}{\bibfnamefont{J.}~\bibnamefont{Meng}} \bibnamefont{and}
  \bibinfo{author}{\bibfnamefont{P.}~\bibnamefont{Ring}},
  \bibinfo{journal}{Phys. Rev. Lett.} \textbf{\bibinfo{volume}{77}},
  \bibinfo{pages}{3963} (\bibinfo{year}{1996}).

\bibitem[{\citenamefont{Meng and Ring}(1998)}]{Meng1998Phys.Rev.Lett.460}
\bibinfo{author}{\bibfnamefont{J.}~\bibnamefont{Meng}} \bibnamefont{and}
  \bibinfo{author}{\bibfnamefont{P.}~\bibnamefont{Ring}},
  \bibinfo{journal}{Phys. Rev. Lett.} \textbf{\bibinfo{volume}{80}},
  \bibinfo{pages}{460} (\bibinfo{year}{1998}).

\bibitem[{\citenamefont{Meng }(1998)}]{Meng1998PRC}
\bibinfo{author}{\bibfnamefont{J.}~\bibnamefont{Meng}},
  \bibinfo{journal}{Phys. Rev. C} \textbf{\bibinfo{volume}{57}},
  \bibinfo{pages}{1229} (\bibinfo{year}{1998}).


\bibitem[{\citenamefont{Meng et~al.}(2002)\citenamefont{Meng, Toki, Zeng,
  Zhang, and Zhou}}]{Meng2002Phys.Rev.C041302}
\bibinfo{author}{\bibfnamefont{J.}~\bibnamefont{Meng}},
  \bibinfo{author}{\bibfnamefont{H.}~\bibnamefont{Toki}},
  \bibinfo{author}{\bibfnamefont{J.~Y.} \bibnamefont{Zeng}},
  \bibinfo{author}{\bibfnamefont{S.~Q.} \bibnamefont{Zhang}}, \bibnamefont{and}
  \bibinfo{author}{\bibfnamefont{S.-G.} \bibnamefont{Zhou}},
  \bibinfo{journal}{Phys. Rev. C} \textbf{\bibinfo{volume}{65}},
  \bibinfo{pages}{041302} (\bibinfo{year}{2002}).

\bibitem[{\citenamefont{Meng}(1998)}]{Meng1998Nucl.Phys.A3}
\bibinfo{author}{\bibfnamefont{J.}~\bibnamefont{Meng}}, \bibinfo{journal}{Nucl.
  Phys. A} \textbf{\bibinfo{volume}{635}}, \bibinfo{pages}{3}
  (\bibinfo{year}{1998}).

\bibitem[{\citenamefont{Zhang et~al.}(2002)\citenamefont{Zhang, Meng, Zhou, and
  Zeng}}]{Zhang2002CPL}
\bibinfo{author}{\bibfnamefont{S.~Q.} \bibnamefont{Zhang}},
  \bibinfo{author}{\bibfnamefont{J.}~\bibnamefont{Meng}},
  \bibinfo{author}{\bibfnamefont{S.-G.} \bibnamefont{Zhou}}, \bibnamefont{and}
  \bibinfo{author}{\bibfnamefont{J.~Y.} \bibnamefont{Zeng}},
  \bibinfo{journal}{Chin. Phys. Lett.} \textbf{\bibinfo{volume}{19}},
  \bibinfo{pages}{312} (\bibinfo{year}{2002}).

\bibitem[{\citenamefont{Zhang et~al.}(2003)\citenamefont{Zhang, Meng, and
  Zhou}}]{zhang2003SIC}
\bibinfo{author}{\bibfnamefont{S.~Q.} \bibnamefont{Zhang}},
  \bibinfo{author}{\bibfnamefont{J.}~\bibnamefont{Meng}}, \bibnamefont{and}
  \bibinfo{author}{\bibfnamefont{S.-G.} \bibnamefont{Zhou}},
  \bibinfo{journal}{Sci. China Ser. G} \textbf{\bibinfo{volume}{46}},
  \bibinfo{pages}{632} (\bibinfo{year}{2003}).

\bibitem[{\citenamefont{Zhou et~al.}(2003)\citenamefont{Zhou, Meng, and
  Ring}}]{Zhou2003Phys.Rev.C034323}
\bibinfo{author}{\bibfnamefont{S.-G.} \bibnamefont{Zhou}},
  \bibinfo{author}{\bibfnamefont{J.}~\bibnamefont{Meng}}, \bibnamefont{and}
  \bibinfo{author}{\bibfnamefont{P.}~\bibnamefont{Ring}},
  \bibinfo{journal}{Phys. Rev. C} \textbf{\bibinfo{volume}{68}},
  \bibinfo{pages}{034323} (\bibinfo{year}{2003}).

\bibitem[{\citenamefont{Zhou et~al.}(2006)\citenamefont{Zhou, Meng, and
  Ring}}]{Zhou2006AIPConf.Proc.90}
\bibinfo{author}{\bibfnamefont{S.-G.} \bibnamefont{Zhou}},
  \bibinfo{author}{\bibfnamefont{J.}~\bibnamefont{Meng}}, \bibnamefont{and}
  \bibinfo{author}{\bibfnamefont{P.}~\bibnamefont{Ring}}, \bibinfo{journal}{AIP
  Conf. Proc.} \textbf{\bibinfo{volume}{865}}, \bibinfo{pages}{90}
  (\bibinfo{year}{2006}).

\bibitem[{\citenamefont{Li et~al.}(2012{\natexlab{a}})\citenamefont{Li, Meng,
  Ring, Zhao, and Zhou}}]{Li2012PhysRevC.85.024312}
\bibinfo{author}{\bibfnamefont{L. L.}~\bibnamefont{Li}},
  \bibinfo{author}{\bibfnamefont{J.}~\bibnamefont{Meng}},
  \bibinfo{author}{\bibfnamefont{P.}~\bibnamefont{Ring}},
  \bibinfo{author}{\bibfnamefont{E.-G.} \bibnamefont{Zhao}}, \bibnamefont{and}
  \bibinfo{author}{\bibfnamefont{S.-G.} \bibnamefont{Zhou}},
  \bibinfo{journal}{Phys. Rev. C} \textbf{\bibinfo{volume}{85}},
  \bibinfo{pages}{024312} (\bibinfo{year}{2012}{\natexlab{a}}).

\bibitem[{\citenamefont{Li et~al.}(2012{\natexlab{b}})\citenamefont{Li, Meng,
  Ring, Zhao, and Zhou}}]{Li2012CPL}
\bibinfo{author}{\bibfnamefont{L. L.}~\bibnamefont{Li}},
  \bibinfo{author}{\bibfnamefont{J.}~\bibnamefont{Meng}},
  \bibinfo{author}{\bibfnamefont{P.}~\bibnamefont{Ring}},
  \bibinfo{author}{\bibfnamefont{E.-G.} \bibnamefont{Zhao}}, \bibnamefont{and}
  \bibinfo{author}{\bibfnamefont{S.-G.} \bibnamefont{Zhou}},
  \bibinfo{journal}{Chin. Phys. Lett.} \textbf{\bibinfo{volume}{29}},
  \bibinfo{pages}{042101} (\bibinfo{year}{2012}{\natexlab{b}}).

\bibitem[{\citenamefont{Chen et~al.}(2012)\citenamefont{Chen, Li, Liang, and
  Meng}}]{Chen2012PhysRevC.85.067301}
\bibinfo{author}{\bibfnamefont{Y.}~\bibnamefont{Chen}},
  \bibinfo{author}{\bibfnamefont{L. L.}~\bibnamefont{Li}},
  \bibinfo{author}{\bibfnamefont{H.}~\bibnamefont{Liang}}, \bibnamefont{and}
  \bibinfo{author}{\bibfnamefont{J.}~\bibnamefont{Meng}},
  \bibinfo{journal}{Phys. Rev. C} \textbf{\bibinfo{volume}{85}},
  \bibinfo{pages}{067301} (\bibinfo{year}{2012}).

\bibitem[{\citenamefont{Bennaceur et~al.}(2000)\citenamefont{Bennaceur,
  Dobaczewski, and Ploszajczak}}]{Bennaceur2000PLB}
\bibinfo{author}{\bibfnamefont{K.}~\bibnamefont{Bennaceur}},
  \bibinfo{author}{\bibfnamefont{J.}~\bibnamefont{Dobaczewski}},
  \bibnamefont{and}
  \bibinfo{author}{\bibfnamefont{M.}~\bibnamefont{Ploszajczak}},
  \bibinfo{journal}{Phys. Lett. B} \textbf{\bibinfo{volume}{496}},
  \bibinfo{pages}{154 } (\bibinfo{year}{2000}).

\bibitem[{\citenamefont{Bertsch et~al.}(1989)\citenamefont{Bertsch, Brown, and
  Sagawa}}]{Bertsch1989_PRC39-1154}
\bibinfo{author}{\bibfnamefont{G.~F.} \bibnamefont{Bertsch}},
  \bibinfo{author}{\bibfnamefont{B.~A.} \bibnamefont{Brown}}, \bibnamefont{and}
  \bibinfo{author}{\bibfnamefont{H.}~\bibnamefont{Sagawa}},
  \bibinfo{journal}{Phys. Rev. C} \textbf{\bibinfo{volume}{39}},
  \bibinfo{pages}{1154} (\bibinfo{year}{1989}).

\bibitem[{\citenamefont{Sagawa}(1992)}]{Sagawa1992_PLB286-7}
\bibinfo{author}{\bibfnamefont{H.}~\bibnamefont{Sagawa}},
  \bibinfo{journal}{Phys. Lett. B} \textbf{\bibinfo{volume}{286}},
  \bibinfo{pages}{7 } (\bibinfo{year}{1992}).

\bibitem[{\citenamefont{Zhu et~al.}(1994)\citenamefont{Zhu, Shen, Cai, and
  Ma}}]{ZHU-ZY1994_PLB328-1}
\bibinfo{author}{\bibfnamefont{Z.~Y.} \bibnamefont{Zhu}},
  \bibinfo{author}{\bibfnamefont{W.~Q.} \bibnamefont{Shen}},
  \bibinfo{author}{\bibfnamefont{Y.~H.} \bibnamefont{Cai}}, \bibnamefont{and}
  \bibinfo{author}{\bibfnamefont{Y. G.}~\bibnamefont{Ma}}, \bibinfo{journal}{Phys.
  Lett. B} \textbf{\bibinfo{volume}{328}}, \bibinfo{pages}{1 }
  (\bibinfo{year}{1994}).

\bibitem[{\citenamefont{Kucharek and Ring}(1991)}]{Kucharek1991_ZPA339-23}
\bibinfo{author}{\bibfnamefont{H.}~\bibnamefont{Kucharek}} \bibnamefont{and}
  \bibinfo{author}{\bibfnamefont{P.}~\bibnamefont{Ring}}, \bibinfo{journal}{Z.
  Phys. A} \textbf{\bibinfo{volume}{339}}, \bibinfo{pages}{23}
  (\bibinfo{year}{1991}).

\bibitem[{\citenamefont{Long et~al.}(2004)\citenamefont{Long, Meng, Giai, and
  Zhou}}]{Long2004Phys.Rev.C034319}
\bibinfo{author}{\bibfnamefont{W. H.}~\bibnamefont{Long}},
  \bibinfo{author}{\bibfnamefont{J.}~\bibnamefont{Meng}},
  \bibinfo{author}{\bibfnamefont{N.~Van} \bibnamefont{Giai}}, \bibnamefont{and}
  \bibinfo{author}{\bibfnamefont{S.-G.} \bibnamefont{Zhou}},
  \bibinfo{journal}{Phys. Rev. C} \textbf{\bibinfo{volume}{69}},
  \bibinfo{pages}{034319} (\bibinfo{year}{2004}).

\bibitem[{\citenamefont{Meng et~al.}(1997)\citenamefont{Meng, P{\"o}schl, and
  Ring}}]{Meng1997_ZPA358-123}
\bibinfo{author}{\bibfnamefont{J.}~\bibnamefont{Meng}},
  \bibinfo{author}{\bibfnamefont{W.}~\bibnamefont{P{\"o}schl}},
  \bibnamefont{and} \bibinfo{author}{\bibfnamefont{P.}~\bibnamefont{Ring}},
  \bibinfo{journal}{Z. Phys. A} \textbf{\bibinfo{volume}{358}},
  \bibinfo{pages}{123} (\bibinfo{year}{1997}).

\bibitem[{\citenamefont{P\"{o}schl et~al.}(1997)\citenamefont{P\"{o}schl,
  Vretenar, Lalazissis, and Ring}}]{Poschl1997_PRL79-3841}
\bibinfo{author}{\bibfnamefont{W.}~\bibnamefont{P\"{o}schl}},
  \bibinfo{author}{\bibfnamefont{D.}~\bibnamefont{Vretenar}},
  \bibinfo{author}{\bibfnamefont{G.~A.} \bibnamefont{Lalazissis}},
  \bibnamefont{and} \bibinfo{author}{\bibfnamefont{P.}~\bibnamefont{Ring}},
  \bibinfo{journal}{Phys. Rev. Lett.} \textbf{\bibinfo{volume}{79}},
  \bibinfo{pages}{3841} (\bibinfo{year}{1997}).

\bibitem[{\citenamefont{Meng et~al.}(1998)\citenamefont{Meng, Tanihata, and
  Yamaji}}]{Meng1998_PLB419-1}
\bibinfo{author}{\bibfnamefont{J.}~\bibnamefont{Meng}},
  \bibinfo{author}{\bibfnamefont{I.}~\bibnamefont{Tanihata}}, \bibnamefont{and}
  \bibinfo{author}{\bibfnamefont{S.}~\bibnamefont{Yamaji}},
  \bibinfo{journal}{Phys. Lett. B} \textbf{\bibinfo{volume}{419}},
   \bibinfo{pages}{1} (\bibinfo{year}{1998}).

\bibitem[{\citenamefont{Vretenar et~al.}(1998)\citenamefont{Vretenar, P\"oschl,
  Lalazissis, and Ring}}]{Vretenar1998_PRC57-1060}
\bibinfo{author}{\bibfnamefont{D.}~\bibnamefont{Vretenar}},
  \bibinfo{author}{\bibfnamefont{W.}~\bibnamefont{P\"oschl}},
  \bibinfo{author}{\bibfnamefont{G.~A.} \bibnamefont{Lalazissis}},
  \bibnamefont{and} \bibinfo{author}{\bibfnamefont{P.}~\bibnamefont{Ring}},
  \bibinfo{journal}{Phys. Rev. C} \textbf{\bibinfo{volume}{57}},
  \bibinfo{pages}{1060} (\bibinfo{year}{1998}).

\bibitem[{\citenamefont{Stoitsov et~al.}(1998)\citenamefont{Stoitsov, Ring,
  Vretenar, and Lalazissis}}]{Stoitsov1998_PRC58-2086}
\bibinfo{author}{\bibfnamefont{M.}~\bibnamefont{Stoitsov}},
  \bibinfo{author}{\bibfnamefont{P.}~\bibnamefont{Ring}},
  \bibinfo{author}{\bibfnamefont{D.}~\bibnamefont{Vretenar}}, \bibnamefont{and}
  \bibinfo{author}{\bibfnamefont{G.~A.} \bibnamefont{Lalazissis}},
  \bibinfo{journal}{Phys. Rev. C} \textbf{\bibinfo{volume}{58}},
  \bibinfo{pages}{2086} (\bibinfo{year}{1998}).

\bibitem[{\citenamefont{Meng}(1999)}]{Meng1999_NPA654-702c}
\bibinfo{author}{\bibfnamefont{J.}~\bibnamefont{Meng}}, \bibinfo{journal}{Nucl.
  Phys. A} \textbf{\bibinfo{volume}{654}}, \bibinfo{pages}{702 }
  (\bibinfo{year}{1999}).

\bibitem[{\citenamefont{Meng et~al.}(2003)\citenamefont{Meng, L\"{u}, Zhang,
  and Zhou}}]{Meng2003_NPA722-366c}
\bibinfo{author}{\bibfnamefont{J.}~\bibnamefont{Meng}},
  \bibinfo{author}{\bibfnamefont{H.~F.} \bibnamefont{L\"{u}}},
  \bibinfo{author}{\bibfnamefont{S.~Q.} \bibnamefont{Zhang}}, \bibnamefont{and}
  \bibinfo{author}{\bibfnamefont{S.-G.} \bibnamefont{Zhou}},
  \bibinfo{journal}{Nucl. Phys. A} \textbf{\bibinfo{volume}{722}},
  \bibinfo{pages}{366} (\bibinfo{year}{2003}).

\bibitem[{\citenamefont{Long et~al.}(2010)\citenamefont{Long, Ring, Meng,
  Van~Giai, and Bertulani}}]{Long2010_PRC81-031302}
\bibinfo{author}{\bibfnamefont{W.~H.} \bibnamefont{Long}},
  \bibinfo{author}{\bibfnamefont{P.}~\bibnamefont{Ring}},
  \bibinfo{author}{\bibfnamefont{J.}~\bibnamefont{Meng}},
  \bibinfo{author}{\bibfnamefont{N.}~\bibnamefont{Van~Giai}}, \bibnamefont{and}
  \bibinfo{author}{\bibfnamefont{C.~A.} \bibnamefont{Bertulani}},
  \bibinfo{journal}{Phys. Rev. C} \textbf{\bibinfo{volume}{81}},
  \bibinfo{pages}{031302} (\bibinfo{year}{2010}).

\bibitem[{\citenamefont{Zhou et~al.}(2010)\citenamefont{Zhou, Meng, Ring, and
  Zhao}}]{ZHOU-SG2010_PRC82-011301}
\bibinfo{author}{\bibfnamefont{S.-G.} \bibnamefont{Zhou}},
  \bibinfo{author}{\bibfnamefont{J.}~\bibnamefont{Meng}},
  \bibinfo{author}{\bibfnamefont{P.}~\bibnamefont{Ring}}, \bibnamefont{and}
  \bibinfo{author}{\bibfnamefont{E.-G.} \bibnamefont{Zhao}},
  \bibinfo{journal}{Phys. Rev. C} \textbf{\bibinfo{volume}{82}},
  \bibinfo{pages}{011301} (\bibinfo{year}{2010}).

\bibitem[{\citenamefont{Zhou et~al.}(2011)\citenamefont{Zhou, Meng, Ring, and
  Zhao}}]{ZHOU-SG2011_JPCS312-092067}
\bibinfo{author}{\bibfnamefont{S.-G.} \bibnamefont{Zhou}},
  \bibinfo{author}{\bibfnamefont{J.}~\bibnamefont{Meng}},
  \bibinfo{author}{\bibfnamefont{P.}~\bibnamefont{Ring}}, \bibnamefont{and}
  \bibinfo{author}{\bibfnamefont{E.-G.} \bibnamefont{Zhao}},
  \bibinfo{journal}{J. Phys. Conf. Ser.} \textbf{\bibinfo{volume}{312}},
  \bibinfo{pages}{092067} (\bibinfo{year}{2011}).

\bibitem[{\citenamefont{Meng et~al.}(2006)\citenamefont{Meng, Toki, Zhou,
  Zhang, Long, and Geng}}]{Meng2006Prog.Part.Nucl.Phys.470}
\bibinfo{author}{\bibfnamefont{J.}~\bibnamefont{Meng}},
  \bibinfo{author}{\bibfnamefont{H.}~\bibnamefont{Toki}},
  \bibinfo{author}{\bibfnamefont{S.-G.} \bibnamefont{Zhou}},
  \bibinfo{author}{\bibfnamefont{S.~Q.} \bibnamefont{Zhang}},
  \bibinfo{author}{\bibfnamefont{W.~H.} \bibnamefont{Long}}, \bibnamefont{and}
  \bibinfo{author}{\bibfnamefont{L.~S.} \bibnamefont{Geng}},
  \bibinfo{journal}{Prog. Part. Nucl. Phys.} \textbf{\bibinfo{volume}{57}},
  \bibinfo{pages}{470} (\bibinfo{year}{2006}).

\bibitem[{\citenamefont{Grasso et~al.}(2001)\citenamefont{Grasso, Sandulescu,
  Van~Giai, and Liotta}}]{Grasso2001_PRC64-064321}
\bibinfo{author}{\bibfnamefont{M.}~\bibnamefont{Grasso}},
  \bibinfo{author}{\bibfnamefont{N.}~\bibnamefont{Sandulescu}},
  \bibinfo{author}{\bibfnamefont{N.}~\bibnamefont{Van~Giai}}, \bibnamefont{and}
  \bibinfo{author}{\bibfnamefont{R.~J.} \bibnamefont{Liotta}},
  \bibinfo{journal}{Phys. Rev. C} \textbf{\bibinfo{volume}{64}},
  \bibinfo{pages}{064321} (\bibinfo{year}{2001}).

  \bibitem[{\citenamefont{Zhang et~al.}(2001)\citenamefont{Zhang, Matsuo,
  and Meng}}]{zhang2011PRC}
\bibinfo{author}{\bibfnamefont{Y.}~\bibnamefont{Zhang}},
  \bibinfo{author}{\bibfnamefont{M.}~\bibnamefont{Matsuo}}, \bibnamefont{and}
  \bibinfo{author}{\bibfnamefont{J.} \bibnamefont{Meng}},
  \bibinfo{journal}{Phys. Rev. C} \textbf{\bibinfo{volume}{83}},
  \bibinfo{pages}{054301} (\bibinfo{year}{2011}).

\bibitem[{\citenamefont{Zhang et~al.}(2010)\citenamefont{Zhang, Liang, and
  Meng}}]{Zhang-Ying2010_IJMPE19-55}
\bibinfo{author}{\bibfnamefont{Y.}~\bibnamefont{Zhang}},
  \bibinfo{author}{\bibfnamefont{H. Z.}~\bibnamefont{Liang}}, \bibnamefont{and}
  \bibinfo{author}{\bibfnamefont{J.}~\bibnamefont{Meng}},
  \bibinfo{journal}{Int. J. Mod. Phys. E} \textbf{\bibinfo{volume}{19}},
  \bibinfo{pages}{55} (\bibinfo{year}{2010}).

\bibitem[{\citenamefont{Ring and Schuck}(1980)}]{Ring1980}
\bibinfo{author}{\bibfnamefont{P.}~\bibnamefont{Ring}} \bibnamefont{and}
  \bibinfo{author}{\bibfnamefont{P.}~\bibnamefont{Schuck}},
  \emph{\bibinfo{title}{The Nuclear Many-Body Problem}}
  (\bibinfo{publisher}{Springer-Verlag}, \bibinfo{address}{Berlin},
  \bibinfo{year}{1980}).

\end{thebibliography}

\end{CJK*}
\end{document}